# Accuracy and Precision of Random Walk with Barrier Model Fitting: Simulations and Applications in Head and Neck Cancers

Jiaren Zou[a]
*Department of Radiation Oncology, University of Michigan, Ann Arbor, MI 48109, USA*
Yue Cao
*Department of Radiation Oncology, University of Michigan, Ann Arbor, MI 48109, USA*
*Department of Radiology, University of Michigan, Ann Arbor, MI 48109, USA*
*Department of Biomedical Engineering, University of Michigan, Ann Arbor, MI 48109, USA*

**Abstract**

Promising results have been reported in quantifying microstructural parameters (free diffusivity, cell size, and membrane permeability) in head and neck cancers (HNCs) using time-dependent diffusion MRI fitted to Random Walk with Barrier Model (RWBM). However, model fitting remains challenging due to limited number of measurements, low signal-to-noise ratio and complex nonlinear biophysical model. In this work, we comprehensively investigated and elucidated the dependence of RWBM fitting performance on tissue property, data acquisition and processing, and provided insights on improving data acquisition and model fitting. We numerically evaluated the accuracy and precision of RWBM fitting using non-linear least squares as a function of model parameters, noise levels, and maximum effective diffusion times over a wide range of microstructural parameter values. We then elucidated these results by examining the model's degeneracy and fitting landscape. *In vivo* fitting results on patients with HNCs were analyzed and were interpreted using the numerical results. We observed that free diffusivity estimates were accurate and precise over a wide range of microstructural parameters, whereas the accuracy and precision of surface-to-volume ratio and membrane permeability depended on the values of the parameters. In addition, a maximum effective diffusion time of $200\ ms$ provided the lowest bias and variance in membrane permeability estimations. By fitting to the short-time limit expression, we observed that the variance of parameter estimation was reduced, and that a more accurate and precise estimation of membrane permeability was achieved. *In vivo* fitting results were consistent with the numerical findings. In conclusion, this work provides a comprehensive analysis of RWBM fitting in clinical settings and has the potential to guide further optimizations of data acquisition and model fitting methods.

Keywords: diffusion MRI, tissue microstructure, model fitting, head and neck cancers

[a] Author to whom correspondence should be addressed. Electronic mail: jiarenz@med.umich.edu.

# 1 Introduction

Diffusion MRI (dMRI) is highly sensitive to hindrance and/or restriction to the free diffusion of water molecules in tissues[1]. This sensitivity enables dMRI to provide valuable insights into various tumor microstructural properties, including cell density, water molecule diffusivity, cell size, and membrane permeability[2,3]. Therefore, it finds numerous applications in cancer diagnosis, prognosis and treatment response assessment[4–7].

There are two major ways of interpreting diffusion measurements. One is to use signal representations to obtain a phenomenological characterization of diffusion of water molecules[8]. For example, cumulant expansion represents the logarithm of the diffusion signal as a Taylor expansion in the diffusion weighting parameter (b-value). The commonly used apparent diffusion coefficient (ADC) is the first order coefficient of the cumulant expansion. These methods, while providing sensitive indices of microstructure, are not specific. For example, water-exchange and cell size can induce conflicting effects on ADC, making it difficult to quantify specific microstructural properties and their changes.

Another approach is to fit analytical models relating time-dependent diffusion signals with microstructure parameters[2] to disentangle effects of different microstructure properties and provide specific biomarkers of tumor microstructure. A number of methods have been developed including POMACE (pulsed and oscillating gradient MRI for assessment of cell size and extracellular space)[9], IMPULSED (imaging microstructural parameters using limited spectrally edited diffusion)[10], and VERDICT (vascular, extracellular, and restricted diffusion for cytometry in tumors)[11]. These methods have demonstrated promising results in the diagnosis and/or treatment response of breast cancer[12–14] and prostate cancer[15–20]. Most existing methods assume impermeable cell membrane, which simplifies the model but introduces bias[21]. In addition, cell membrane permeability changes, e.g. due to tumor cell response to treatment[22], cannot be measured by these models, which could serve as a responsive biomarker to therapy. Recently, RWBM (random walk with barrier model)[23] with modeling of membrane permeability showed promising results in extracting microstructural parameters from time-dependent dMRI of head and neck squamous cell carcinomas (HNSCCs)[24]. The cell size metric was consistent with pathological findings. In addition, these parameters demonstrated significant correlation with tumor staging and exhibited significant changes in response to chemoradiation therapy (CRT). Despite the potential of model-based approaches, the model fitting is a challenge for clinical data due to the interplay of sparse data sampling, low signal-to-noise ratio (SNR) and increasing number of model parameter to fit[2,25,26]. The low accuracy and precision of parameter estimates can negatively impact the spatial resolution of microstructure mapping and the statistical power of associated analyses. Low precision also restricts current evaluations to mean parameter values in tumors, overlooking the heterogeneity within tumor volumes, which may further provide clinically useful information of treatment response[27]. A systematic characterization and understanding of the accuracy and precision of tumor microstructure parameter estimation is essential for the optimization of protocol and model fitting.

In this work, we comprehensively investigated and elucidated behaviors of RWBM fitting by non-linear least squares (NLLS) from clinically feasible diffusion measurements. Through simulation studies, we investigated the accuracy and precision of RWBM fitting with variations in parameter values (cell size and membrane permeability), noise levels, and effective diffusion times. These results were elucidated by examining the degeneracy and fitting landscape of RWBM. By *in vivo* studies, we evaluated the distributions and correlations of the fitted microstructure parameters in HNSCCs and compared them with numerical findings.

## 2 Methods

### 2.1 Theory

RWBM describes tissue microstructure by randomly placed and oriented permeable membranes [23]. It characterizes dispersive diffusivity of water molecules by

$$\mathcal{D}(\omega) = \frac{D_0}{1 + \zeta + 2z_\omega(1 - z_\omega)\left[\sqrt{1 + \frac{\zeta}{(1 - z_\omega)^2}} - 1\right]}, z_\omega = i\sqrt{i\omega\tau}, \quad (1)$$

where $D_0$ is the free diffusion coefficient; $\zeta = (S/V)l/d$ is the effective "volume fraction" occupied by the membrane; and $\tau = l^2/D_0 = D_0/(2\kappa)^2$ is the characteristic time scale for a single membrane. $l = D_0/2\kappa$ is the effective thickness. $\kappa$ is the membrane permeability. $S/V$ is the membrane surface-to-volume ratio. d is the dimension of the diffusion space. The relationship between the dispersive diffusivity $\mathcal{D}(\omega)$ and the cumulative diffusion coefficient $D(t)$ is established via[26,28]

$$D(t) \equiv \frac{\langle(r(t) - r(0))^2\rangle}{2t} = \frac{1}{t}\int_{-\infty}^{\infty} \frac{d\omega}{2\pi} \frac{\mathcal{D}(\omega)e^{-i\omega t}}{(-i\omega)^2}, \tag{2}$$

where $r(t)$ is the location of a water molecule at time $t$. Therefore, $D(t)$ can be evaluated at different microstructural parameter combinations through numerical integration as in Fieremans et al.[26]. Equation 2 can thus be used for diffusion signal simulation and model fitting[26].

In the short-time limit, i.e., $t \ll \tau_D$, where $\tau_D = \bar{a}^2/2D_0$ is the diffusion time across typical cell size, and $\bar{a}$ is the typical cell size $\bar{a} \simeq 2Vd/S$, the cumulative diffusion coefficient has a geometry-independent expression:

$$D(t) \cong D_0\left[1 - \frac{S}{Vd}\left(\frac{4}{3}\sqrt{\frac{D_0 t}{\pi}} - \kappa t\right)\right], \tag{3}$$

which has been used for model fitting in microstructural parameter mapping in HNSCCs[24]. Although promising results have been demonstrated, it is not clear whether the short-limit condition is satisfied in HNSCCs, which may introduce bias into the parameter estimates. Therefore, it is important to investigate the model fitting behavior of Equation 3 across a wide range of microstructural parameters.

The diffusion signal is $S(b,t) = S_0 e^{-bD(t)}$, where $S_0$ is the unweighted signal amplitude and b is the b-value. The commonly used NLLS fitting amounts to minimizing an objective function in the space of three RWBM parameters $D_0$, $V/S$, and $\kappa$:

$$F = \frac{1}{N_{b,t}} \sum_i^{N_{b,t}} (S_i - S(b_i, t_i))^2, \tag{4}$$

where $N_{b,t}$ is total number of diffusion measurements $S_i$. $V/S$ was used instead of $S/V$ for easier interpretation as a metric of cell size.

## 2.2 Numerical simulation study

We investigated model fitting at biophysically plausible parameter combinations (BP-PCs) of $D_0 = \{2, 3\}\ \mu m^2/ms$, $V/S = [1.2: 0.19: 5]\mu m$, and $\kappa = [0.03: 0.0035: 0.1]\mu m/ms$, as well as $D_0 = 2.0\mu m^2/ms$, $V/S = 4.0\mu m$, and $\kappa = 0.06\mu m/ms$ which were representative microstructure parameters in HNSCCs[24] (HNSCC-PC). The diffusion signals were simulated using general RWBM (Equation 1) with b-values and effective diffusion times of a clinically feasible sequence (Table 1)[24]. The effective diffusion times $t$ used in this work were calculated following Fordham et al.[29]. Specifically, $t$ was approximated as $t = 2t_0$ as in Cao et al.[24], where $t_0 = \Delta - \delta/3$ for pulse gradient spin echo (PGSE) and $t_0 = 1/3f$ for sinusoid oscillating gradient spin echo (OGSE). Here, $\Delta$ denotes the time interval between dephasing and rephasing gradient waveforms, $\delta$ is the width of the pulse gradient waveform, and $f$ is the frequency of the oscillating gradients. Complex Gaussian noise was added to each synthetic diffusion signal, with the real and imaginary components following independent Gaussian distributions with the same SNR. The SNR was defined as the ratio of the b=0 signal magnitude to the standard deviation of the real or imaginary component of the noise. The magnitude signals were fitted by NLLS using both general RWBM (G-NLLS, equation 2) and STL-RWBM (STL-NLLS, equation 3) with constraints $10^{-4} \leq D_0 \leq 3.3 \mu m^2/ms$, $10^{-4} \leq V/S \leq 5 \mu m$, and $10^{-4} \leq \kappa \leq 0.1 \mu m/ms$ implemented by curve_fit function of SciPy package [30].

**Table 1.** Effective diffusion time t defined by Fordham et al.[29], along with corresponding b-values, repetition times (TRs), and echo times (TEs) used for diffusion measurements in simulation and *in vivo* studies.

|      | t [ms] | b-values [$ms/\mu m^2$] | TR [ms] | TE [ms] |
|------|--------|-------------------------|---------|---------|
| OGSE | 13.3   | 0.05, 0.1, 0.2, 0.32    |         |         |
|      | 19.0   | 0.05, 0.3, 0.6, 0.9     | 3100    | 89      |
| PGSE | 34.7   | 0.05, 0.3, 0.5, 0.75, 0.9 |       |         |
|      | 56.3   | 0.05, 0.3, 0.5, 0.75, 1 |         |         |

PGSE: pulse gradient spin echo; OGSE: oscillating gradient spin echo.

### 2.2.1 Impact of $V/S$ and $\kappa$

We investigated the dependence of biases and standard deviations (SDs) on $V/S$ and $\kappa$. We generated a dictionary of RWBM parameters-signal pairs at BP-PCs. Complex Gaussian noise with SNR of 50 was added with 100 noise realizations. Relative bias ($|\bar{x} - x|/x$) and relative SD ($\sqrt{\frac{1}{N_s}\sum_j^{N_s}(\bar{x} - x_j)^2}/x$) were evaluated, where $x$ is the ground truth parameter, $\bar{x}$ is the average estimate across noise realizations, $x_j$ is the estimate for the j$^{th}$ noise realization, $N_s$ is the total number of noise realizations. Bias and SD were evaluated for each parameter combination and reported as functions of $V/S$ and $\kappa$ at two $D_0$ values: 2 and 3 $\mu m^2/ms$.

### 2.2.2 Sensitivity to noise

The dependence of biases and SDs of the model parameters on SNR levels was also investigated. The diffusion signal was generated at HNSCC-PC, and complex Gaussian noise was added at SNR levels of 20, 40, 60, 80, and 100, with 100 noise realizations per noise level. In addition to relative bias and relative SD, normalized root mean squared error (NRMSE) ($\sqrt{\frac{1}{N_S}\sum_j^{N_S}(x-x_j)^2}/x$) were evaluated for each SNR.

### 2.2.3 Sensitivity to maximum diffusion time

We investigated the impact of maximum diffusion times on the parameter estimation. We generated diffusion signals at HNSCC-PC using ten q-t sampling schemes with different maximum diffusion times $t_{max}$ of 20, 40, 60, 80, 100, 200, 400, 600, 800, 1000 ms. For each sampling scheme, five diffusion times were uniformly sampled between 10 ms and the corresponding maximum value. For each diffusion time, five b-values were uniformly sampled between 0.05 $ms/\mu m^2$ and the corresponding typical maximum b-value that can be achieved on a clinical scanner. OGSE was used to measure diffusion signals with diffusion times lower than 20 ms and with a maximum b-value of 0.5 $ms/\mu m^2$. For diffusion times between 20 and 80 ms, PGSE was for simulation with a maximum b-value of 1 $ms/\mu m^2$. Diffusion times longer than 80 ms can be achieved by stimulated echo acquisition mode (STEAM)[31] with a maximum b-value of 1 $ms/\mu m^2$. To account for the reduced signal from stimulated echoes and T$_1$ relaxation, the STEAM signal was modeled as $\frac{1}{2}S_0 e^{-T_M/T_1} e^{-bD(t)}$, where $T_M = t - T_E/2$ represents the mixing time and T$_1$ was assumed to be 1500 ms[32]. Additionally, T$_1$ was included as a fitted tissue parameter for measurements involving STEAM sequences. Complex Gaussian noise with SNR of 50 was added with 100 noise realizations. Relative bias and relative SD were evaluated for each q-t sampling scheme.

### 2.2.4 Degeneracy in RWBM fitting

Degeneracy in model fitting means that multiple sets of parameters can describe the acquired signal equally well, leading to ambiguity in model fitting[33]. Evaluating this degeneracy is essential for understanding the bias and variance in model fitting. In this experiment, we investigated the degeneracy in RWBM parameter estimation. First, a noiseless MR signal was simulated based on the RWBM for the HNSCC-PC. Next, complex Gaussian noise with an SNR of 50 was added, generating 2,500 noise realizations. Subsequently, model fitting was performed on each noisy realization using random initial guesses drawn from uniform distributions: $D_0 \in [0, 3.3]\ \mu m^2/ms$, $V/S \in [0, 5]\mu m$, and $\kappa \in [0, 0.1]\ \mu m/ms$. The noiseless signal was also fitted 2,500 times, with each fit using a different random initial guess sampled from the same uniform distributions. The histograms of the estimated parameters were reported and visually analyzed.

### 2.2.5 RWBM fitting landscape

The fitting landscape was evaluated for the HNSCC-PC by calculating the value of the objective function (equation 4) at all parameter combinations of $D_0 = [1.35: 0.02: 3.3]\ \mu m^2/ms$, $V/S = [1.2: 0.04: 5]\ \mu m$, and $\kappa = [0.03: 0.0007: 0.1]\ \mu m/ms$. The landscape was evaluated in the absence of noise, as well as in the presence of complex Gaussian noise (SNR = 50).

## 2.3 *In vivo* study in HNSCCs

Ten patients with HNSCCs were enrolled on the institutional review board approved protocol. All patients provided written consent. The images were acquired using a ~10-minute acquisition protocol (Table 1) on the 3T clinical scanner (MAGNETOM Skyra, Siemens Healthcare, Erlangen, Germany) using a research echo-planar imaging pulse sequence that included two OGSE scans (f = 50 and 35 Hz) and two PGSE scans ($\Delta/\delta = 39.0\ ms/32.7\ ms$ and $\Delta/\delta = 22.0\ ms/13.9\ ms$). A sinusoid trapezoid waveform with four half lobes was chosen for all OGSEs to achieve higher b-values. Diffusion weighting in three orthogonal directions was applied with three averages for each b-value. Parallel imaging factor of 3 was used for short echo trains and reduction of susceptibility-induced geometric distortion. 18 slices were acquired for each patient with in-plane resolution of 2.3×2.3 mm, and slice thickness of 5 mm with a 1.5 mm gap for consideration of cross-talking between adjacent slices. All diffusion-weighted images were corrected for $B_0$ inhomogeneity-induced geometric distortion and signal loss. In each image session, additional post-Gadolinium (Gd) $T_1$-weighted images with one standard dose (0.1 mmol/kg) of MultiHance were acquired using a two-dimensional (2D) fast spin echo sequence with an in-plane resolution of ~1×1 mm and 3 mm slice thickness with 10% gap. Patients were scanned using individual radiation therapy (RT) immobilized devices, with 5-point face masks that were tightened to a flat tabletop to constrain head and neck motion.

Diffusion-weighted images were fit voxel-by-voxel to both general and STL RWBM by NLLS. The histograms of microstructure parameters within the primary tumors of all patients were compared across fitting methods. Two-dimensional histograms of parameters were used to evaluate correlations between parameters. Example fitted parameter maps overlaid on post-Gd $T_1$-weighted images were visually evaluated. The experiments were repeated on 3×3 Gaussian filtered diffusion-weighted images to investigate the impact of the commonly applied Gaussian filtering.

## 3 Results

### 3.1 Numerical simulation study

#### 3.1.1 Impact of $V/S$ and $\kappa$

We investigated how the biases and SDs of estimated $D_0$, $V/S$, and $\kappa$ depend on the ground truth values of $V/S$ and $\kappa$, in which two representative $D_0$ values of 2 and 3 $\mu m^2/ms$ were used (Figure 1). For both G-NLLS and STL-NLLS, biases and SDs of $D_0$ estimates have weak dependence on $V/S$ and $\kappa$ values. The $D_0$ biases and SDs were less than 15% for most $V/S$ and $\kappa$ values. However, biases and SDs of estimated $V/S$ and $\kappa$ strongly depended on their corresponding ground truth values.

For $V/S$ estimation, the bias by G-NLLS depended on both $V/S$ and $\kappa$ values, whereas the bias by STL-NLLS depended solely on $V/S$. The bias by G-NLLS was less than 35% except for $V/S < 3\ \mu m$ with $\kappa > 0.05\ \mu m/ms$, while the bias by STL-NLLS was less than 35% for $V/S > 3\mu m$. The SDs by both methods depended mostly on $V/S$ values and little on $D_0$ and $\kappa$, with lower SDs for larger $V/S$.

For $\kappa$ estimation, bias by both methods showed dependence on both $V/S$ and $\kappa$ values. The estimates at small $V/S$ with large $\kappa$ and large $V/S$ with small $\kappa$ had the largest bias by both methods. The SDs by both methods showed similar patterns and depended on both $V/S$ and $\kappa$ values, showing low SDs at large $V/S$ and large $\kappa$.

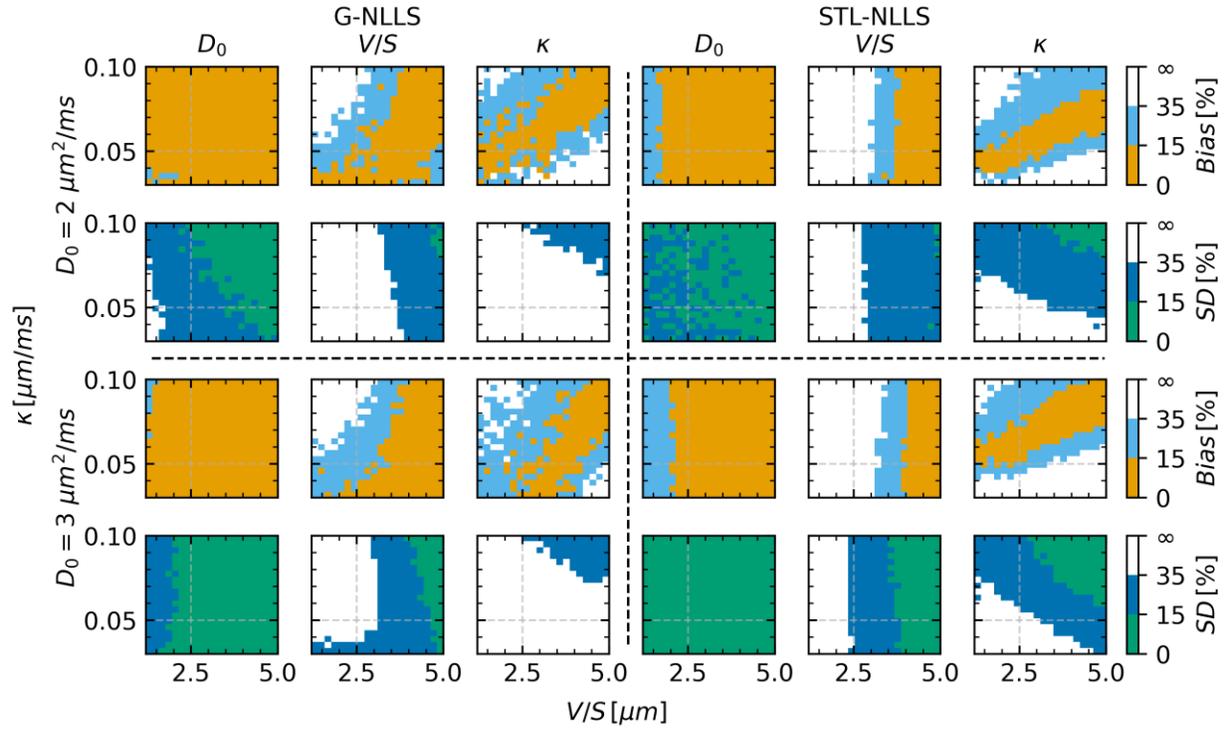

**Figure 1.** Bias and SD of microstructural parameter estimation as a function of $V/S$ and $\kappa$ for G-NLLS and STL-NLLS. Two $D_0$ values (2 and 3 $\mu m^2/ms$) were evaluated. The SNR was 50.

### 3.1.2 Sensitivity to noise

We investigated the sensitivity of model fitting to noise at HNSCC-PC (Figure 2). For $D_0$ estimation, G-NLLS consistently outperformed STL-NLLS in terms of NRMSE, bias and SD for $SNR \leq 80$. Interestingly, STL-NLLS outperformed G-NLLS in terms of NRMSE for both $V/S$ and $\kappa$ estimations across all SNR levels. Although STL-NLLS exhibited a higher bias in $V/S$ estimation, this was offset by its lower SD, leading to an overall lower NRMSE. In $\kappa$ estimation, STL-NLLS outperformed G-NLLS in all evaluation metrics when $SNR \geq 40$.

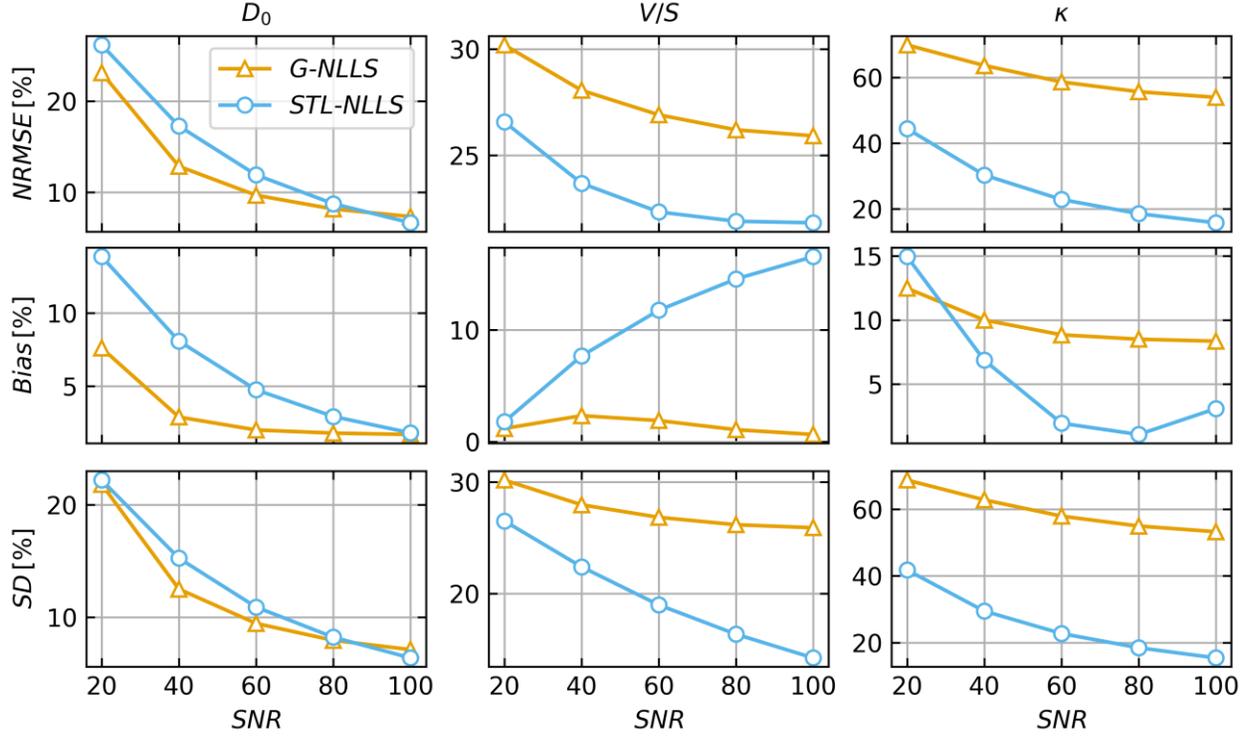

**Figure 2.** NRMSE (first row), bias (second row), and SD (third row) of RWBM microstructural parameter estimation at HNSCC-PC by G-NLLS and STL-NLLS across different SNRs.

### 3.1.3 Sensitivity to maximum diffusion time

We investigated how the maximum diffusion time affects the bias and SD of parameter estimation, in which the same number of measurements were maintained to ensure a fair comparison (Figure 3). For G-NLLS, the biases and SDs of $D_0$ and $V/S$ estimates were stable across measurements with different $t_{max}$. However, the bias of $\kappa$ first decreased as $t_{max}$ increased from $20\ ms$ to $200\ ms$ and then increased for $t_{max} > 200\ ms$. The SD of estimated $\kappa$ showed a decreasing trend with $t_{max}$ except at $t_{max} = 100 ms$ where STEAM was first included in the measurements.

For STL-NLLS, the bias and SD of estimated $D_0$ remained stable across $t_{max}$. The biases of $V/S$ and $\kappa$ estimates heavily depended on $t_{max}$. The bias of $V/S$ estimates increased with $t_{max}$ until $t_{max} = 200 ms$ where the bias reached a plateau, whereas the bias of estimated $\kappa$ first decreased until $t_{max} = 60 ms$ and then increased with $t_{max}$. Both SDs of $V/S$ and $\kappa$ estimates decreased with $t_{max}$.

By comparing G-NLLS and STL-NLLS, we observed that both methods yielded similar levels of bias and SD of $D_0$ estimation. G-NLLS achieved a lower bias in the estimated $V/S$, whereas its

SD was slightly higher than that of STL-NLLS. Notably, STL-NLLS estimation of $\kappa$ at $t_{max}\sim 60ms$ achieved both lower bias and lower SD compared to G-NLLS.

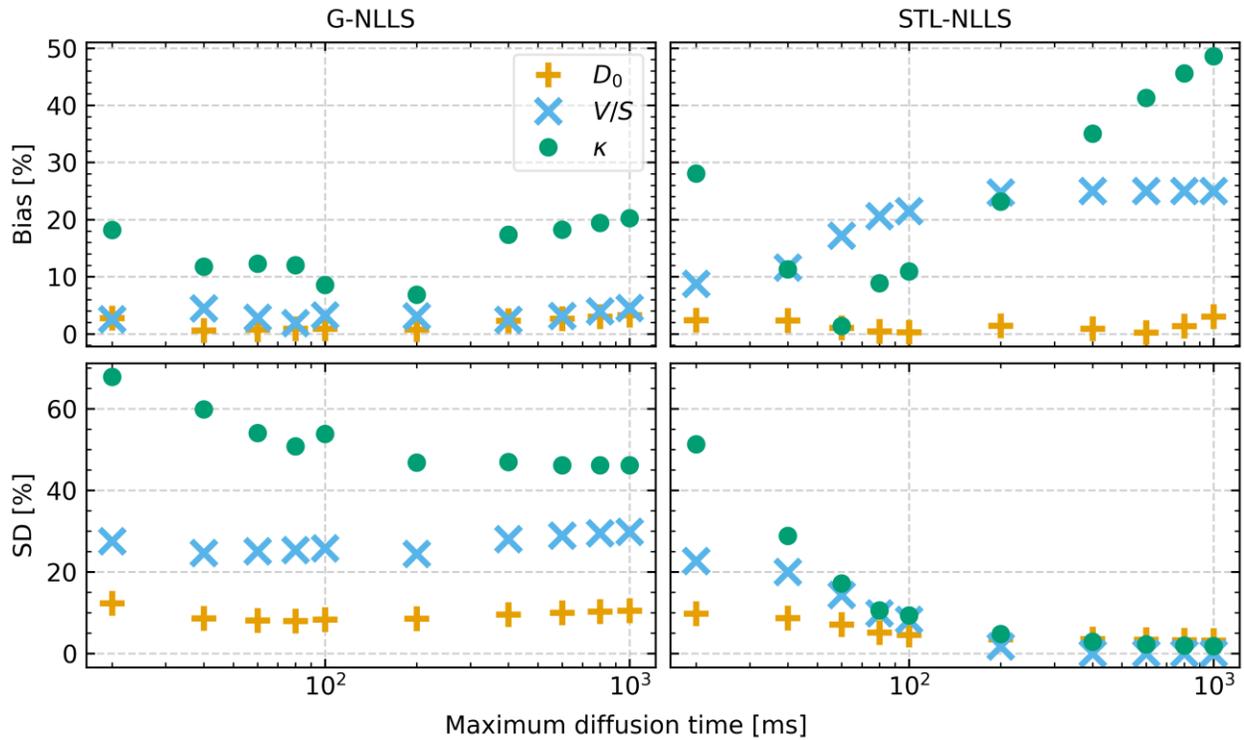

**Figure 3.** Parameter estimation bias (first row) and SD (second row) as a function of maximum diffusion time for G-NLLS (first column) and STL-NLLS (second column).

### 3.1.4 Degeneracy of RWBM fitting

The degeneracy of RWBM fitting was investigated by fitting 2500 noise realizations of diffusion signals of HNSCC-PC using random initializations (Figure 4). For G-NLLS, when noise was absent, the parameter estimation accurately recovered the ground truth values, which indicates that there is no intrinsic degeneracy in the RWBM fitting for HNSCC-PC using diffusion measurements described in this study. This contrasts with multicompartment models, which inherently exhibit degeneracies[33]. However, when noise (SNR=50) was introduced to the diffusion signals, the distribution of fitted parameters became substantially broader. Estimations reached parameter constraints for all parameters. STL-NLLS estimates of $V/S$ and $\kappa$ with noise exhibited narrower distributions around the ground truth values than G-NLLS, even though the estimates were biased without noise.

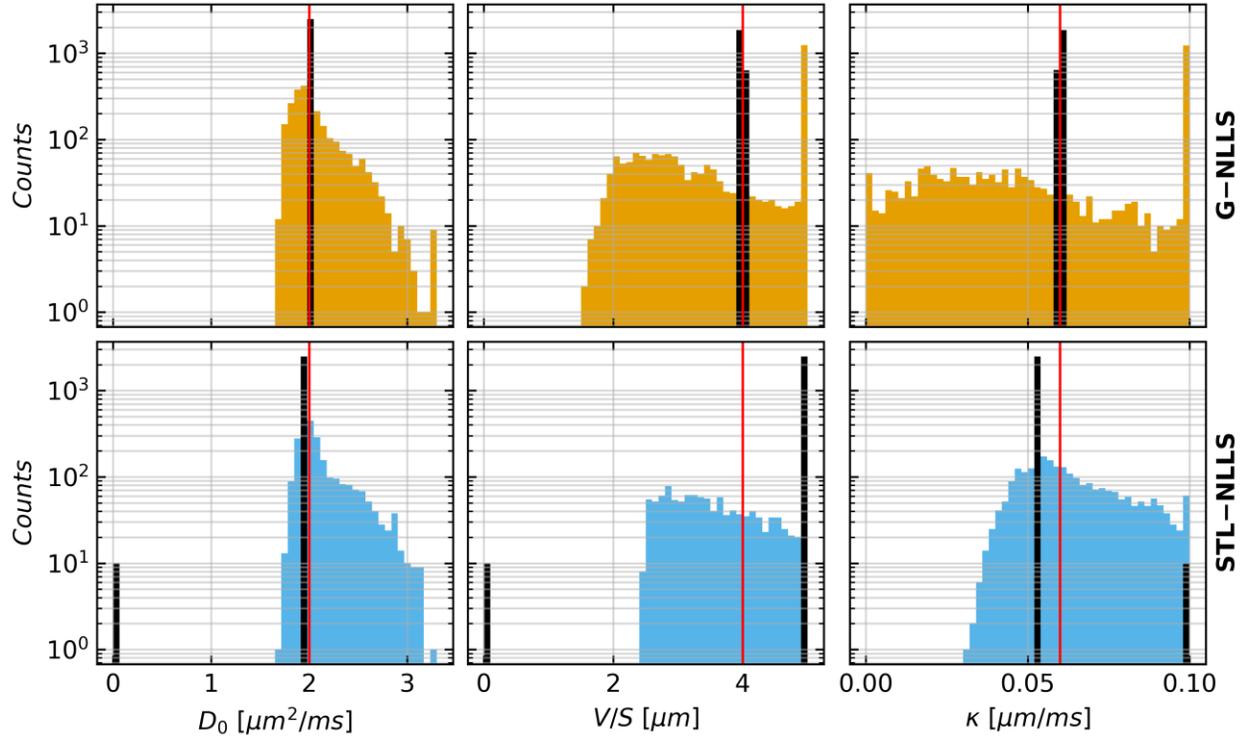

**Figure 4.** Histograms of fitted RWBM parameters for HNSCC-PC at SNR = ∞ (black histograms), SNR = 50 using G-NLLS (yellow histograms), and SNR = 50 using STL-NLLS (blue histograms). Red line: ground truth ($D_0 = 2.0 \mu m^2/ms$, $V/S = 4.0 \mu m$, and $\kappa = 0.06 \mu m/ms$).

### 3.1.5 RWBM fitting landscape

We investigated the fitting landscape of RWBM for HNSCC-PC by visualizing the 2D landscapes corresponding to different $\kappa$ values (Figure 5). For G-NLLS, in the noiseless case, no spurious minimum disconnected from the global minimum was observed. This is in contrast with multi-compartment models, which exhibit inherent degeneracies with the number of solutions equal to the number of compartments[33]. Compared with the noiseless case, the fitting landscape when SNR=50 was flatter where a wide range of parameter combinations on a pipe can have similar objective value. This may lead to widely varying fitting results and apparent correlations between microstructure parameters. Note that the landscape extended beyond the parameter constraints used for fitting, which may lead to estimates hitting the boundaries of the constraints. For STL-NLLS, the fitting landscape was less flat compared to G-NLLS, which may contribute to the smaller SDs of STL-NLLS fitting.

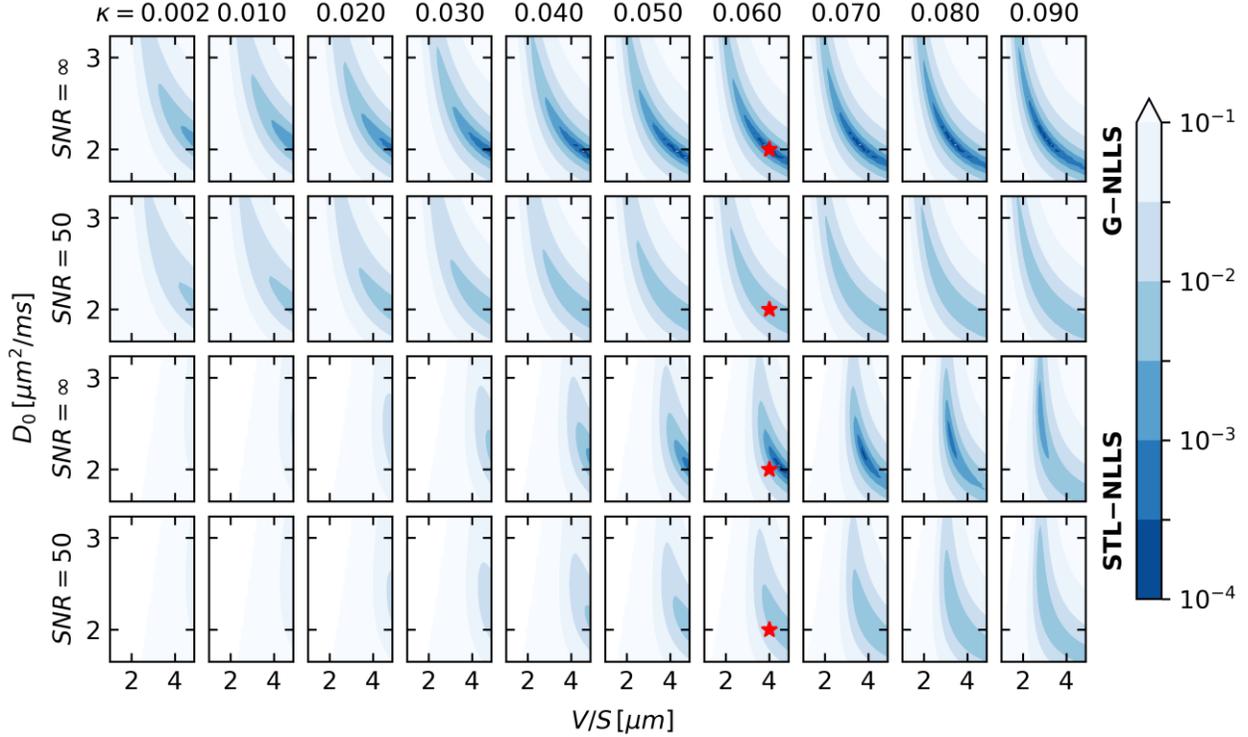

**Figure 5.** The fitting landscapes showing the value of the objective (equation 4) as a function of $V/S$ and $D_0$ at ten $\kappa$ values. The landscapes for G-NLLS and STL-NLLS are shown in the first two and last two rows, respectively. Results at $SNR = \infty$ and SNR=50 are displayed for each fitting method in separate rows. The red star indicates the ground truth parameter combination ($D_0 = 2.0 \mu m^2/ms$, $V/S = 4.0 \mu m$, and $\kappa = 0.06 \mu m/ms$). $\kappa$ is in the unit of $\mu m/ms$.

### 3.2 In vivo study in HNCs

G-NLLS and STL-NLLS fitting methods were investigated on *in vivo* primary HNSCCs of ten patients. The histograms of the fitted parameters of all tumor voxels (Figure 6) showed that the G-NLLS estimates had a large number of voxels hitting the boundaries of the parameter constraints for all three parameters, with $D_0$ and $V/S$ hitting the upper bound and $\kappa$ hitting both lower and upper bounds. Note that this result is consistent with the findings of the degeneracy study where $D_0$ and $V/S$ estimates had a clear peak at the upper bound and $\kappa$ had peaks at both lower and upper bounds with noise present (Figure 4). STL-NLLS reduced the number of voxels hitting the boundaries compared to G-NLLS. With Gaussian filtering, the number of voxels reaching the upper bound of $D_0$ decreased for both G-NLLS and STL-NLLS, but not for the other two parameters.

The mean of the estimated microstructural parameters in the tumors from G-NLLS and STL-NLLS are $D_0 = 2.0 \pm 0.3 \mu m^2/ms$, $V/S = 3.8 \pm 0.6 \mu m$, $\kappa = 0.07 \pm 0.01 \mu m/ms$, and $D_0 = 1.8 \pm 0.1 \mu m^2/ms$, $V/S = 2.7 \pm 0.1 \mu m$, $\kappa = 0.057 \pm 0.004 \mu m/ms$, respectively. The SD was calculated across ten patients. The G-NLLS estimated $V/S$ is consistent with the pathological findings in a previous work[24] which corresponds to $V/S = 3.6 - 4.0 \ \mu m$. The STL-NLLS yields biased $V/S$, whereas it reduces SDs in all parameters.

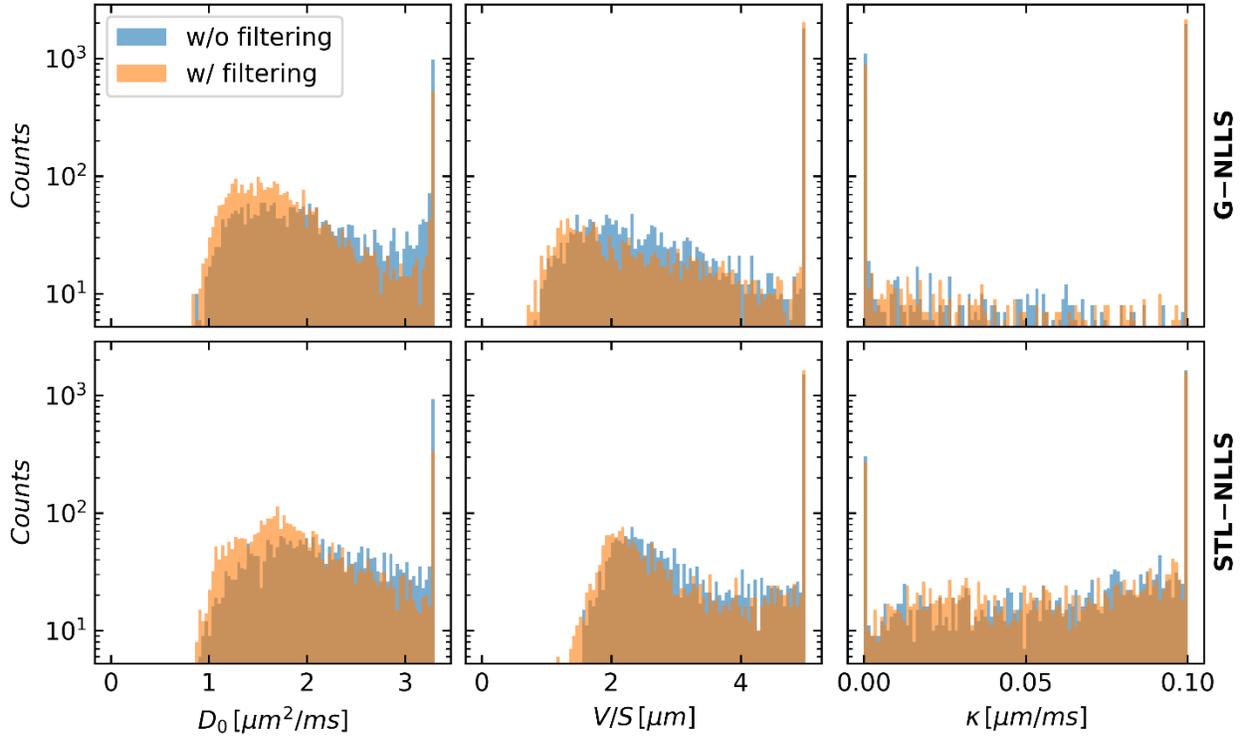

**Figure 6.** Histogram of RWBM parameter estimates in primary HNSCCs of ten patients using G-NLLS (first row), and STL-NLLS (second row) with (blue histogram) and without (orange histogram) Gaussian filtering.

We also plotted 2D histograms of the fitted parameters in all tumor voxels to investigate the correlation between parameters (Figure 7). In G-NLLS, a negative correlation was observed between $D_0$ and $V/S$. Similarly, in STL-NLLS, a negative correlation was identified between $V/S$ and $\kappa$.

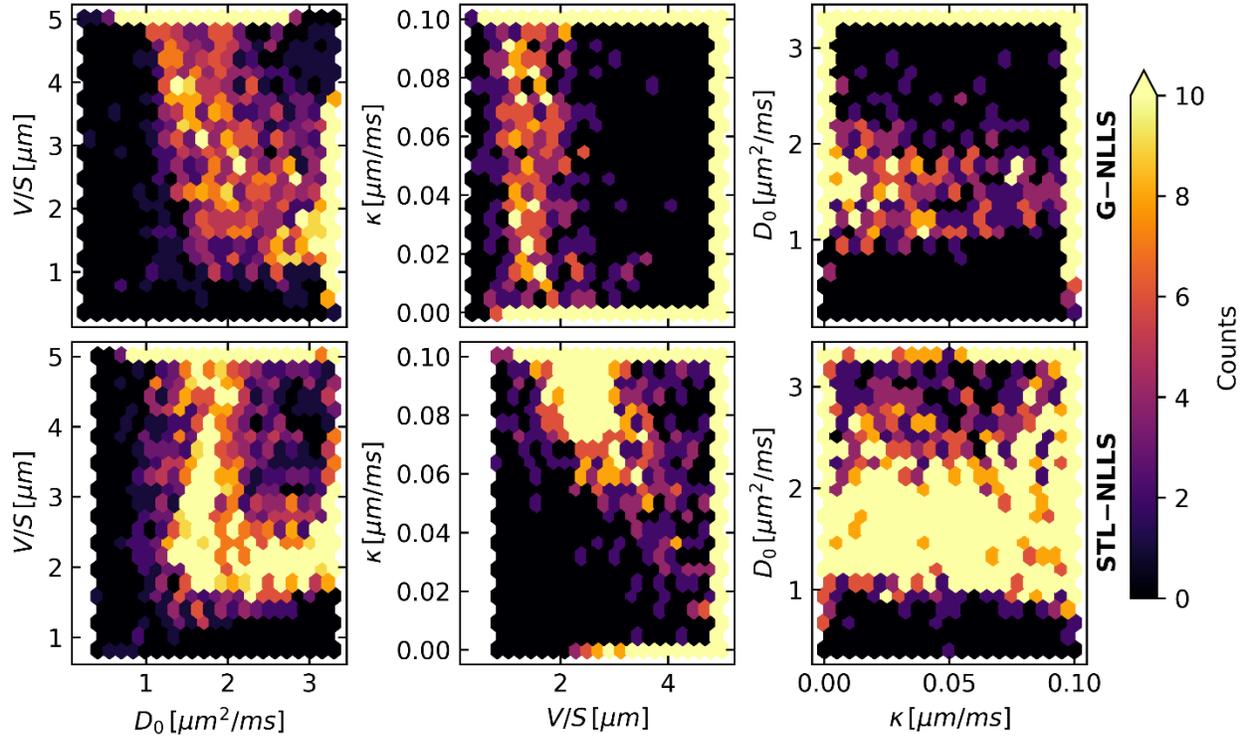

**Figure 7.** 2D histogram of RWBM parameter estimates in primary HNSCCs of ten patients using G-NLLS (first row), and STL-NLLS (second row). 3×3 Gaussian filtering was applied before model fitting.

Example parameter maps are shown in Figure 8, in which the $D_0$ maps estimated by G-NLLS and STL-NLLS were highly consistent and showed similar spatial heterogeneity. Differences between the two methods appeared in the $V/S$ and $\kappa$ maps. In some patients, G-NLLS provided unrealistic maps with many voxels hitting the boundaries of the parameter constraints (e.g., $V/S$ map of patient 3 and $\kappa$ map of patient 6), whereas STL-NLLS reduced the number of these voxels. This observation is consistent with the simulation study.

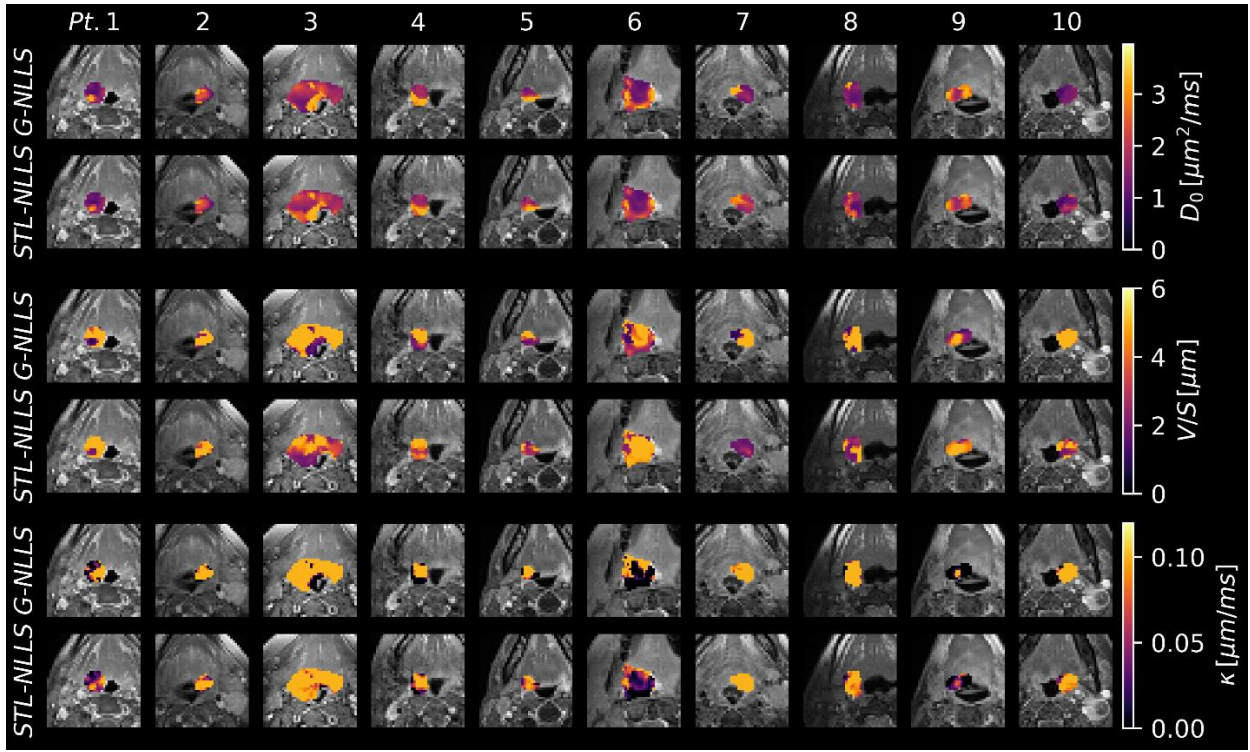

**Figure 8.** Example fitted microstructure parameter maps of ten patients with HNSCCs by both G-NLLS and STL-NLLS overlaid on post-Gd $T_1$ weighted images. The mid tumor slice was shown for each patient. 3×3 Gaussian filtering was applied before model fitting.

## 4 Discussion

In this work, we performed a comprehensive evaluation of RWBM fitting characteristics, focusing on its sensitivity to microstructural parameters and noise, as well as the degeneracy and landscape of model fitting, using a clinically feasible diffusion protocol. By simulation studies, we observed that the biases and SDs of the estimations depend on the underlying microstructural parameters, maximum effective diffusion time, and fitting models (general RWBM vs. STL expression). The degeneracy and fitting landscape studies partly explained these dependences and revealed parameter inter-connections with noise present. The findings from simulation studies are consistent with model fitting results on *in vivo* data of HNSCCs. These findings can guide future development of sequence design and model fitting methods.

Compared to other diffusion models for tumor microstructure, such as IMPULSED and POMACE, the fitting accuracy and precision of cell size metrics are comparable. In a simulation study, IMPULSED model slightly overestimates cell diameter in the presence of transcytolemmal water exchange and noise[21]. A large variance of cell diameter estimates was observed when cell diameter was in the range of 10-15 $\mu m$. The estimates also reached upper bound of the constraint for cell diameter (30 $\mu m$) as observed in this work. Note that they used nine b-values between 0 and 2 $ms/\mu m^2$ for both PGSE and OGSE measurements which is a much wider range than what was used in this study. On simulated GL261 glioma signals (SNR=120), POMACE cell radius estimates also present large variations[2] with a range of 5 $\mu m$ while the ground truth is 4.8 $\mu m$. In that experiment, ten OGSE steps from 60 to 225 Hz and four PGSE steps (6/9/16/31 ms) were used which is also a much richer set of measurements requiring much longer acquisition time that cannot be achieved in current clinical settings.

Dependences of biases and SDs on the ground truth parameters are revealed (Figure 1). In regions of low $V/S$ and high $\kappa$, due to the limit of the shortest diffusion time or the highest OGSE frequency achievable on clinical scanners, $V/S$ and $\kappa$ cannot be well characterized. For large $V/S$ and low $\kappa$, insufficient water molecules have reached or crossed the cell membrane at the longest measured diffusion time, resulting in increased bias in $V/S$ and $\kappa$. Although signal reduction and T$_1$ relaxation decrease the SNR of STEAM measurements, extending $t_{max}$ beyond 80 ms reduces both the bias and SD of $\kappa$ estimates for G-NLLS (Figure 3). However, further increasing $t_{max}$ beyond 200 ms increases the bias, possibly due to the too much lower SNR. Similar results were reported in a simulation study based on random permeable cylinders[34]. The low bias estimation of $\kappa$ by STL-NLLS when $t_{max} = 60 ms$ could be related to the fitting landscape (Figure 5) where it extends over a wide range of $\kappa$ for G-NLLS whereas the landscape of STL-NLLS is narrower around the ground truth $\kappa$.

From the fitting landscape of RWBM at HNSCC-PC (Figure 5), we can see that the objective values at the upper bounds of $D_0$ and $V/S$ lie on the same level set as the global minimum. However, this is not the case for the lower bounds of $D_0$ and $V/S$, which explains the observation that the estimates of $D_0$ and $V/S$ tend to reach the upper parameter constraints but not the lower ones. For $\kappa$, the objective values were on a similar scale across different $\kappa$ values for G-NLLS, allowing the estimates to reach both the lower and upper bounds. In contrast, for STL-NLLS, the objective values are higher at smaller $\kappa$, reducing the number of voxels that reach the lower bound.

The apparent correlations between $D_0$ and $V/S$, as well as between $V/S$ and $\kappa$, align with the fitting landscape. The negative correlation between $D_0$ and $V/S$ can be intuitively understood. For most of the diffusion times used in this study, $t \sim \tau_D$. In this time regime, a sufficient number of water molecules reach the cell membrane and are partially restricted by the membrane. By increasing $V/S$ while keeping $D_0$ and $\kappa$ unchanged, the water molecules would travel longer distances before hitting the cell membrane at the time of measurement, reducing diffusion signal and thus increasing the objective. By decreasing $D_0$, this effect can be balanced and thereby maintaining similar diffusion signals. Similarly, a smaller $\kappa$ can increase the restriction and limit the number of water molecules traveling outside the cells, thereby balancing the signal loss caused by an increased $V/S$. These trade-offs between $D_0$ and $V/S$, as well as between $V/S$ and $\kappa$ extend to unphysical parameter combinations, especially when noise is present, resulting in estimates that reach parameter constraints.

This study suggests that optimizing acquisition protocols has the potential to enhance the accuracy and precision of RWBM fitting. Optimization based on Cramer-Rao lower bound or neural networks will be interesting directions to explore. The study also suggests that the choice of fitting model depends on the parameter range and metrics of interest. For instance, STL-NLLS can be beneficial when fitting stability is preferred, even when the short-time limit condition is not met. This study focused solely on conventional NLLS fitting. Incorporating prior knowledge into model fitting using neural networks or Bayesian methods holds the potential to further enhance parameter estimation performance[16,25,35] and is a promising direction for future investigation.

## 5 Conclusions

In conclusion, this work provides a comprehensive characterization of RWBM fitting under clinically feasible diffusion measurements. Dependences of fitting bias and SD on microstructural parameters, effective diffusion times, and fitting models were investigated and elucidated. These dependencies can guide further optimization of acquisition protocols and model fitting methods.